\documentclass{WileyMSP-template}
\usepackage{amsmath} 
\usepackage{physics} 

\begin{document}

\pagestyle{fancy}

\title{Problem of nonlinear conductivity within relaxation time approximation in noncentrosymmetric insulators}

\maketitle


\author{Ibuki Terada}
\author{Sota Kitamura}
\author{Hiroshi Watanabe}
\author{Hiroaki Ikeda}


\begin{affiliations}
I. Terada, H. Ikeda\\
Department of Physical Science, Ritsumeikan University, Shiga 525-8577, Japan\\
Email Address: rp0068xi@ed.ritsumei.ac.jp

S. Kitamura\\
Department of Applied Physics, The University of Tokyo, Hongo, Tokyo, 113-8656, Japan

H. Watanabe\\
Department of Liberal Arts and Basic Sciences, College of Industrial Technology, Nihon University, Chiba 275-8576, Japan
\end{affiliations}


\begin{abstract}
With the recent advancements in laser technology, there has been increasing interest in nonlinear and nonperturbative phenomena such as nonreciprocal transport, the nonlinear Hall effect, and nonlinear optical responses. 
When analyzing the nonequilibrium steady state, the relaxation time approximation (RTA) in the quantum kinetic equation has been widely used. 
However, recent studies have highlighted problems with the use of RTA that require careful consideration.
In a study published in Phys. Rev. B, {\bf 109}, L180302 (2024), we revealed that the RTA has a flaw in predicting finite linear conductivity even for insulators under weak electric fields, and improved the RTA based on the Redfield equation.
In this paper, we further extend our approach to nonlinear responses. 
This approach provides a simple alternative to RTA and is expected to be useful for the study of nonlinear and nonequilibrium phenomena.
\end{abstract}


\section{Introduction}
The responses of quantum materials to external fields has been actively studied not only in the linear regime but also in the nonlinear regime, with various phenomena such as nonreciprocal transport~\cite{NatCommun.1.3740,PhysRevLett.87.236602,NatPhys.6.578}, the nonlinear Hall effect~\cite{NatRevPhys.11.744,PhysRevLett.115.216806,NatMat.4.324}, and nonlinear optical responses~\cite{boyd2003,bloembergen1996}. Most of these responses are understood through evaluation of higher-order response tensors with respect to external fields, primarily based on perturbation theory. However, in recent years, with the development of laser technologies, there has been growing interest in nonperturbative effects induced by strong external fields, which cannot be explained by conventional perturbation theory. These nonperturbative effects are particularly recognized as important in phenomena such as nonreciprocal transport~\cite{PhysRevB.102.245141}, photocurrent generation~\cite{Nature.7675.224,10.21468/SciPostPhys.11.4.075} and high-harmonic generation~\cite{NatPhys.2.138,NatPhoto.2.119}.

Calculating the nonequilibrium steady state in nonperturbative regimes is generally a challenging problem. Typically, Landau-Zener tunneling --- quantum tunneling between different levels under a strong field --- has been extensively studied in isolated systems~\cite{IVAKHNENKO20231}. 
However, research on the dynamics of electrons in solids has been limited to recent studies focused on specific materials, such as graphene.
In these studies, the response to strong external fields is often analyzed numerically using the quantum kinetic equation in momentum $k$-space~\cite{PhysRevB.99.214302,PhysRevResearch.2.043408,PhysRevB.110.035303}, given by
\begin{equation}\label{QKE}
\frac{d\rho_k(t)}{dt}=-i\bigg[\mathcal H_k(t),\rho_k(t)\bigg]+\mathcal{D}_k(\rho_k(t)),
\end{equation}
where $\rho_k(t)$ is the density matrix and $\mathcal{H}_k(t)$ is the Hamiltonian involving the external field. The second term represents the dissipation term, which drives the electron dynamics towards a nonequilibrium steady state. 
In practice, this dissipation term is often handled using a phenomenological relaxation time approximation (RTA), as follows.
\begin{equation}\label{RTA}
\mathcal D(\rho_k(t))=-\frac{\rho_k(t)-\rho^{\rm eq}_{k}}{\tau}
\end{equation}
with $\rho_k^{\rm eq}$ representing the density matrix in the equilibrium state. $\tau^{-1}$ denotes the damping rate.

The RTA has long been used as a convenient method for describing nonequilibrium steady states. However, in recent years, its limitations and problems of this approach have been explicitly pointed out ~\cite{PhysRevB.97.235446,PhysRevB.103.195133}.  We have recently revealed that the RTA has fatal flaws in current responses of insulators under DC fields~\cite{PhysRevB.109.L180302}. 
We revealed that the RTA has a flaw in predicting finite linear conductivity even for insulators under weak electric fields and proposed the Dynamical Phase Approximation (DPA) method, which improves upon the RTA using the Redfield equation. Here, we demonstrate that similar issues arise in the nonlinear response regime and investigate a formalism to accurately describe second-order responses using the DPA.
Given the increasing focus on nonlinear and nonperturbative response phenomena, identifying and remedying the problems inherent in frequently used RTA is essential for accurately understanding the nonlinear and nonequilibrium phenomena.

\section{Nonlinear conductivity within RTA}
To examine the problem of RTA in the nonlinear response regime, we begin by expanding the density matrix within RTA up to second order in the DC electric field $E$. Following Ref.~\cite{PhysRevB.109.L180302}, we consider a two-band system with a finite gap $\Delta_k=\varepsilon_{k+}-\varepsilon_{k-}>0$ at all $k$-points, where $\varepsilon_{k\pm}$ are the two eigenenergies.
The DC electric field is introduced via the Peierls substitution, $k \rightarrow k - Et$ (hereafter $\hbar = e = 1$).
The eigenenergy at time $t$, $\varepsilon_{k\alpha}(t)$, is obtained from the following equation:
$\mathcal{H}_k(t)\ket{u_{k\alpha}(t)} = \varepsilon_{k\alpha}(t) \ket{u_{k\alpha}(t)}$,
where $\mathcal{H}_k(t)=\mathcal{H}_{k-Et}$ represents the Hamiltonian at the $k$-point, written in the $2\times2$ matrix form.

First, we introduce the adiabatic wavefunction (snapshot basis):
\begin{subequations}
\begin{align}
& \ket{\Phi_{k\alpha}(t)} = e^{-i\theta_{k\alpha}(t)}\ket{u_{k\alpha}(t)}, \\
& \theta_{k\alpha}(t) = \int_{t_0}^t [\varepsilon_{k\alpha}(t')+EA_{k,\alpha\alpha}(t')] dt',
\end{align}
\end{subequations}
where $A_{k,\alpha\beta}(t) = i \braket{u_{k\alpha}(t)}{\partial_k u_{k\beta}(t)}$ is the Berry connections. 
In this basis, the quantum kinetic equation can be recast as
\begin{equation}
\frac{d\rho_k(t)}{dt} = -i\left[\mathcal{W}_k(t),\rho_k(t)\right]-\frac{\rho_k(t)-\rho_k^{\rm eq}(t)}{\tau},
\end{equation}
where $[\mathcal{W}_k(t)]_{\alpha\beta}=EA_{k,\alpha\beta}(t)e^{i\{\theta_{k\alpha}(t)-\theta_{k\beta}(t)\}}(1 - \delta_{\alpha\beta})$ is the transition dipole matrix elements and $[\rho^{\rm eq}_k(t)]_{\alpha\beta}=f_D(\varepsilon_{k\alpha}(t))\delta_{\alpha\beta}$ with $f_D$ the Fermi-Dirac distribution function. 
For simplicity, following Ref.~\cite{PhysRevB.109.L180302}, we assume that the damping rate $\tau^{-1}$ is band-independent.
By introducing the time evolution operator $U(t,t')=\mathcal{T}\exp[-i\int_{t'}^t\mathcal{W}_k(s)ds]$, the formal solution of Eq. \eqref{QKE} can be expressed as
\begin{equation}
\rho_k(t)=U(t,t_0)\rho_k(t_0)U^\dag(t,t_0)e^{-(t-t_0)/\tau}+\frac{1}{\tau}\int_{t_0}^tU(t,s)\rho_k^{\rm eq}(s)U(s,t)e^{-(t-s)/\tau}ds.
\end{equation}
We set $t_0\rightarrow -\infty$ and $\Delta_k(t_0)\rightarrow\infty$, hereafter. 
In the weak electric field regime, the time evolution operator can be approximated as $[U(t,t')]_{\alpha\beta}\simeq \delta_{\alpha\beta}-i\int_{t'}^{t}[\mathcal W_k(s)]_{\alpha\beta}ds$.
In this case, each component of the density matrix is given by
\begin{subequations}
\begin{align}
&[\rho_k(t)]_{\alpha\alpha}\simeq \tau^{-1}\int_{t_0}^{t}f_D(\varepsilon_{k\alpha}(s))e^{-(t-s)/\tau}ds,\\
&[\rho_k(t)]_{+-}\simeq\tau^{-1}\int_{t_0}^{t}\left(-i\int_{s_2}^{t}[\mathcal W_k(s_1)]_{+-}ds_1\right)\delta f_k(s_2)e^{-(t-s_2)/\tau}ds_2,
\end{align}
\end{subequations}
where $\delta f_k(t)=f_D(\varepsilon_{k-}(t))-f_D(\varepsilon_{k+}(t))$. 
Here, the second-order terms of $[\mathcal W_k(s)]_{\alpha\beta}$ 
are neglected because they do not contribute to the non-diagonal terms that are of primary interest.

Expanding the density matrix up to the second order in $E$, we obtain
\begin{subequations}
\begin{align}
\label{Diagonal_RTA}
&[\rho_k(t)]_{\alpha\alpha}\simeq f_D(\varepsilon_{k\alpha}(t))+E\tau\left.\frac{\partial f_D}{\partial k}\right|_{\varepsilon=\varepsilon_{k\alpha}(t)}+(E\tau)^2\left.\frac{\partial^2 f_D}{\partial k^2}\right|_{\varepsilon=\varepsilon_{k\alpha}(t)},
\\
\label{Offdiagonal_RTA}
\begin{split}
&[\rho_k(t)]_{+-}\simeq -\frac{[\mathcal W_k(t)]_{+-}}{\Delta_k(t)-i\tau^{-1}}\bigg[\bigg\{1-E\frac{z_{k+}(t)+\tau^{-1}\partial_k\ln\Delta_k(t)/\Delta_k(t)}{\Delta_k(t)-i\tau^{-1}}\bigg\}\delta f_k(t)\\
&~~~~~~~~~~~~~~~~~~~~~~~~~~~~~~~~~
+E\tau \frac{\Delta_k(t)-i2\tau^{-1}}{\Delta_k(t)-i\tau^{-1}} \frac{\partial\delta f_{k}(t)}{\partial (k-Et)}\bigg],
\end{split}
\end{align}
\end{subequations}
where $z_{k\alpha}=R_k+i\,\alpha\,\partial_k\ln|A_{k,+-}|$.
$~R_k=A_{k,++}-A_{k,--}-\partial_k\arg A_{k,+-}$ is the geometric quantity called shift vector, which plays an important role in shift current responses~\cite{doi:10.7566/JPSJ.92.072001,PhysRevLett.109.236601,doi:10.1126/sciadv.1501524} and nonreciprocal Landau-Zener tunneling~\cite{10.21468/SciPostPhys.11.4.075,CommunPhys.1.63}. The diagonal component \eqref{Diagonal_RTA} reproduces the result of the Boltzmann equation.

Assuming that the electric current in the weak electric field regime behaves as $J=\sigma^{(1)}E+\sigma^{(2)}E^2$, we aim to evaluate the linear conductivity $\sigma^{(1)}$ and the nonlinear conductivity $\sigma^{(2)}$. Then, we introduce the velocity in the snapshot basis, $v_{k,\alpha\beta}(t) = \mel**{\Phi_{k\alpha}(t)}{\partial_k \mathcal{H}_k(t)}{\Phi_{k\beta}(t)}$, and define the current as
\begin{subequations}
\begin{align}
&J(t)=-e\int\frac{dk}{2\pi}\mathrm{Tr}[v_k(t)\rho_k(t)]=J_{\rm intra}(t)+J_{\rm inter}(t),\\
&J_{\rm intra}(t)=-\sum_{\alpha=\pm}\int\frac{dk}{2\pi}\frac{\partial \varepsilon_{k\alpha}(t)}{\partial k}[\rho_k(t)]_{\alpha\alpha},\\
&J_{\rm inter}(t)=-2\mathrm{Im}\left[\int\frac{dk}{2\pi}\frac{[W_k(t)]_{-+}\Delta_k(t)}{E}[\rho_k(t)]_{+-}\right].
\end{align}
\end{subequations}
Here, the electric current is separated into an intraband contribution $J_{\rm intra}$ and an interband contribution $J_{\rm inter}$. The electric current in the nonequilibrium steady state is obtained from the long-time limit of $J(t)$.

The intraband contribution to the conductivity is given by
\begin{subequations}
\begin{align}
\label{Intra1_RTA}
&\sigma^{(1)}_{\rm intra}=\tau\sum_{\alpha}\int\frac{dk}{2\pi}\frac{\partial \varepsilon_{k\alpha}}{\partial k}\left(-\frac{\partial f_D(\varepsilon_{k\alpha})}{\partial k}\right),\\
\label{Intra2_RTA}
&\sigma^{(2)}_{\rm intra}=-\tau^2\sum_{\alpha}\int\frac{dk}{2\pi}\frac{\partial^2\varepsilon_{k\alpha}(t)}{\partial k^2}\left(-\frac{\partial f_D(\varepsilon_{k\alpha})}{\partial k}\right).
\end{align}
\end{subequations}
Both the linear and nonlinear conductivities reproduce the Drude conductivity. Equation~\eqref{Intra2_RTA} is used, for example, in the analysis of magneto-chiral anisotropy~\cite{PhysRevLett.87.236602,NatPhys.6.578} and spin currents~\cite{PhysRevB.95.224430,PhysRevB.106.085127} in the systems lacking inversion symmetry. On the other hand, the interband contribution to the electrical conductivity is given by
\begin{subequations}
\begin{align}
\label{Inter1_RTA}
&\sigma^{(1)}_{\rm inter}=2\int\frac{dk}{2\pi}\frac{\Delta_k|A_{k,+-}|^2}{\tau(\Delta^2_k+\tau^{-2})}\delta f_k,\\
\label{Inter2_RTA}
\begin{split}
&\sigma^{(2)}_{\rm inter}=-2\int\frac{dk}{2\pi}\bigg[\frac{\delta f_k}{(\Delta_k^2+\tau^{-2})^2}\bigg\{\frac{2\Delta_k^2|A_{k,+-}|^2R_{k}}{\tau}+\frac{\Delta_k(\Delta^2_k-\tau^{-2})\partial_k|A_{k,+-}|^2}{2}+\frac{|A_{k,+-}|^2\partial_k \Delta_k}{\tau^2}\bigg\}\\
&~~~~~~~~~~~~~~~~~~~~~~~~~~~~~~~~~~~~~~~~~~~~~~~~~~~~~~~~~~
-\frac{2\Delta_k|A_{k,+-}|^2}{\tau^2(\Delta_k^2+\tau^{-2})^2}\frac{\partial}{\partial k}\bigg(f_{D}(\varepsilon_{k-})-f_{D}(\varepsilon_{k+})\bigg)\bigg].
\end{split}
\end{align}
\end{subequations}

The linear conductivity $\sigma^{(1)}_{\rm inter/intra}$ reproduces the result in Ref.~\cite{PhysRevB.109.L180302}.
In the insulating case, the intraband conductivity $\sigma^{(1),(2)}_{\rm intra}$ vanishes at $T=0$ as $\frac{\partial f_D(\varepsilon_{k\alpha})}{\partial k}\to 0$, which is consistent with the Boltzmann theory. However, the interband conductivity $\sigma^{(1),(2)}_{\rm inter}$ yields a finite conductivity despite the absence of Fermi surfaces. This result is counterintuitive in understanding insulating behavior in weak electric fields and is considered as an artifact of RTA. In the previous study, we clarified that the unphysical contribution $\sigma^{(1)}_{\rm inter}$ is completely canceled out by properly incorporating the first-order contribution of the electric field $E$ beyond RTA.
The expression in Eq.~\eqref{Inter1_RTA} within the RTA emerges because $[\rho_k(t)]_{+-}\simeq -\frac{W_k(t)}{\Delta_k(t)-i\tau^{-1}}\delta f_k(t)$ holds in low $E$ limit. However, in the exact calculation, this is corrected to $[\rho_k(t)]_{+-}\simeq-\frac{W_k(t )}{\Delta_k(t)}\delta f_k(t)$, which leads to the cancellation of the unphysical $\sigma^{(1)}_{\rm inter}$ proportional to $\tau^{-1}$.
Similarly, the first term proportional to $\tau^{-1}$ in Eq.~\eqref{Inter2_RTA} is an artifact of the RTA and should be correctly canceled out. 
Note that this term contains the shift vector $R_k$, which is problematic only in noncentrosymmetric systems. 
The other terms that do not involve $R_k$ vanish when the system has time-reversal symmetry, i.e., in the absence of magnetization.

In the next section, we show how the DPA in Ref.~\cite{PhysRevB.109.L180302} corrects the fatal flaw of the RTA in electric conductivity up to $E^2$, which leads to unphysical nonreciprocal transport in noncentrosymmetric systems.

\section{Improvements of RTA in nonlinear response regime}
First, let us overview the derivation of the DPA.
To microscopically introduce dissipation, we consider a quantum open system $\tilde H=\sum_k \{\tilde H_k(t)+\tilde  H_{B,k}(t)+\tilde H_{I,k}(t) \}$,
where the tilde symbol denotes the interaction picture, 
a two-band system of interest $\tilde H_k=\sum_{\sigma\sigma'}\mel**{\sigma}{\mathcal H_k(t)}{\sigma'}\tilde{c}^\dag_{k\sigma}(t)\tilde{c}_{kp\sigma'}(t)$, a fermionic bath $\tilde  H_{B,k}=\sum_{p\sigma}\omega_p\tilde{b}^\dag_{kp\sigma}(t)\tilde{b}_{kp\sigma}(t)$ and 
an interaction
$\tilde H_{I,k}=\sum_{p\sigma}V_p\tilde{b}^\dag_{kp\sigma}(t)\tilde{c}_{k\sigma}(t)+h.c.$. $\tilde{c}_{k\sigma}$ and $\tilde{b}_{kp\sigma}$ are the annihilation operators for electron in the two-band system and the thermal bath with wavevector $k$ and pseudospin $\sigma$, respectively. 
Electrons in the thermal bath also have degrees of freedom for mode $\omega_p$. 

We microscopically construct the quantum kinetic equation for this quantum open system using the Redfield equation~\cite{breuer2007},
\begin{equation}\label{RME}
\frac{d\tilde\rho_k(t)}{dt}=-\int_{-\infty}^t\mathrm{Tr}_B[\tilde H_{I,k}(t),[\tilde H_{I,k}(s),\tilde\rho_k(s)\otimes\tilde\rho_B]]ds,
\end{equation}
where $\mathrm{Tr}_B[\cdots]$ denotes the trace over the degrees of freedom of the thermal bath. 
The Redfield equation provides time evolution of the density matrix that describes the Markovian dynamics of a quantum open system weakly coupled to a thermal bath (Born-Markov approximation). 
Next, we impose the broadband condition for the spectral density of the fermionic bath,
$\sum_p \pi |V_p|^2 \delta(\omega-\omega_p)={\frac{1}{2\tau}}$,
and express Eq.~\eqref{RME} in the snapshot basis 
$\tilde\psi_{k\alpha}(t)=\sum_{\sigma}\braket{\Phi_{k\alpha}(t)}{\sigma}\tilde c_{k\sigma}(t)$.
In addition, as a key approximation in the DPA, we linearize the dynamical phase factor of $[\mathcal W_k(t)]_{+-}$ and $\ket{\Phi_{k\alpha}(t)}$ in the integrand of Eq.~\eqref{RME} as
\begin{subequations}
\begin{align}
&[\mathcal W_k(t-s)]_{+-}\sim e^{-i\Delta_k(t) s}[\mathcal W_k(t)]_{+-},\\
&\ket{\Phi_{k\alpha}(t-s)}\sim e^{+i\varepsilon_{k\alpha}(t) s}\ket{\Phi_{k\alpha}(t)}.
\end{align}
\end{subequations}
This approximation allows us to analytically perform the integration in Eq. \eqref{RME} and microscopically determine the dissipation term.
The dissipation term in Eq.~\eqref{QKE}, $\mathcal{D}_k(\rho_k(t))$, is corrected in the DPA. 
Consequently, the unphysical interband linear conductivity in RTA, \eqref{Inter1_RTA}, is completely cancelled out in the DPA. 
Our formalism restores the insulating properties in the linear response regime.

In the previous section, we identified the presence of an unphysical $E^2$ term in the interband conductivity within the RTA. This term is also expected to be corrected in the DPA.
Here, we apply the DPA to the interband conductivity
up to second order in $E$, addressing the issues of the RTA in the nonlinear response regime. 
To this end, let us refine
the approximation for time evolution of $[\mathcal W_k(t)]_{+-}$ to second order in $E$ as follows:
\begin{equation}\label{DPA-w}
[\mathcal W_k(t-s)]_{+-}\simeq  e^{-i\Delta_k(t) s}[\mathcal W_k(t)]_{+-}\bigg[1-iEs z_{k+}(t)\bigg].
\end{equation}
Strictly speaking, Eq. \eqref{DPA-w} does not account for all the $E^2$-terms. However, as we will see later, this approximation is sufficient to cancel the unphysical interband conductivity in RTA.

For simplicity, we focus on insulators with particle-hole symmetry ($\varepsilon_{k+}=-\varepsilon_{k-}$) in the following. Consequently, in this case, the dissipation term in Eq.~\eqref{QKE} can be summarized as follows,
\begin{subequations}
\begin{align}
&\mathcal D_k=\mathcal D^{(0)}_{k}+\mathcal D^{(1)}_{k}+\mathcal D^{(2)}_{k},\\
&[\mathcal D^{(0)}_{k}]_{\alpha\beta}=-\frac{[\rho_k(t)]_{\alpha\beta}-f_D(\varepsilon_{k\alpha}(t))\delta_{\alpha\beta}}{\tau},\\
&[\mathcal D^{(1)}_{k}]_{\alpha\beta}=-\frac{[\mathcal W_k(t)]_{\alpha\beta}\delta f_k(t)}{\tau\Delta_k(t)}\bigg(1-\frac{E}{\Delta_k(t)}z_{k\alpha}(t)\bigg),\\
&[\mathcal D^{(2)}_{k}]_{\alpha\beta}=\alpha\frac{|[\mathcal W_k(t)]_{+-}|^2\delta f_k(t)}{\tau\Delta^2_k(t)}\bigg(1-\frac{E}{\Delta_k(t)}z_{k+}(t)\bigg)\bigg(1-\frac{E}{\Delta_k(t)}z_{k-}(t)\bigg)\delta_{\alpha\beta}.
\end{align}
\end{subequations}
Since we are focusing on insulators here, the terms containing the first-order derivative of $f_D$ are excluded.
$\mathcal D^{(1)}_k$ and $\mathcal D^{(2)}_k$ represent the corrections to the dissipation term $\mathcal D^{(0)}_k$ in the RTA, and the off-diagonal term $\mathcal D^{(1)}_{k}$ is the key correction obtained in the DPA. On the other hand, $\mathcal D^{(2)}_{k}$ plays a role in restoring the CPTP (Completely Positive Trace-Preserving) property.

Within this approximation, the off-diagonal components of the density matrix up to the second order of $E$ are as follows,
\begin{equation}\label{DPA}
[\rho_k(t)]_{+-}\simeq -\frac{W_k(t)\delta f_k(t)}{\Delta_k(t)}\bigg(1-E\,\frac{R_k(t)+i\partial_k\ln|A_{k,+-}|}{\Delta_k(t)}\bigg).
\end{equation}
It should be noted once again that terms involving the derivative of $f_D$ are excluded. Comparison with Eq.~\eqref{Offdiagonal_RTA} reveals that the result in the DPA, Eq.~\eqref{DPA}, does not include $\tau^{-1}$ in the denominator.
As a result, the interband nonlinear conductivity, Eq.~\eqref{Inter2_RTA}, is significantly modified as
\begin{equation}\label{Inter2_DPA}
\sigma^{(2)}_{\rm inter}=-\int\frac{dk}{2\pi}\frac{\partial_k|A_{k,+-}|^2}{\Delta_k}\delta f_k.
\end{equation}
This integral vanishes since $|A_{k,+-}|$
, $\Delta_k(t)$, and $\delta f_k$ are even functions of $k$ in the systems with time-reversal symmetry. Consequently, the unphysical interband conductivity observed in the RTA is absent in the DPA, ensuring insulating behavior in the weak DC electric field regime.

\section{Summary}
In this paper, we demonstrated that the critical flaw in the RTA appears not only in the linear conductivity under a DC electric field but also in nonlinear conductivity. We addressed this issue by evaluating higher-order terms in the electric field, based on the DPA we previously proposed. Our results suggest that this flaw in the RTA could extend to even higher-order responses. However, since nonperturbative currents driven by tunneling effects dominate in the strong electric field regime, the issue observed in weak electric fields is unlikely to pose a major concern. Our findings offer a straightforward alternative to the RTA and are expected to be valuable in the study of nonlinear and non-equilibrium phenomena.

\medskip

\medskip
\textbf{Acknowledgements} \par 
We are grateful to Y. Michishita, K. Takasan, A. Oguri, M. Sato, N. Kawakami, T. Morimoto and T. Oka for valuable comments. This work was supported by KAKENHI Grants No. 19H01842, No. 19H05825. SK is supported by JSPS KAKENHI Grant No. 20K14407.
This study is supported by ``Society for the Advancement of Science and Technology at Ritsumeikan".

\medskip

%
\bibliographystyle{MSP}
\bibliography{reference}


\end{document}